\documentclass[aps,prl,twocolumn]{revtex4-1}

\usepackage{graphicx}
\usepackage{amsmath,amssymb}
\usepackage{bm}
\usepackage{physics}
\usepackage{hyperref}

\begin{document}

\title{Quantum Interference Breaks Bias Symmetry at Extended Superconducting Interfaces}

\author{Vishal Tripathi}
\author{Goutam Sheet}
\email{goutam@iisermohali.ac.in}
\affiliation{Department of Physical Sciences, Indian Institute of Science Education and Research (IISER) Mohali, Sector 81, S. A. S. Nagar, Manauli, PO 140306, India}

\date{\today}

\begin{abstract}
Particle-hole symmetry of the Bogoliubov-de~Gennes Hamiltonian is widely assumed to enforce bias-symmetric transport at superconducting interfaces. We show that this expectation fails generically for interfaces with finite spatial extent due to quantum interference. Using a tight-binding scattering formalism that preserves exact particle-hole symmetry, we demonstrate that propagation through an extended interface causes electrons and holes to accumulate unequal phases, leading to intrinsic bias-asymmetric conductance. The interface thereby acts as an effective Andreev interferometer with characteristic damped oscillations arising from coherent multiple reflections within the barrier. While the asymmetry originates from normal-state interference, its bias dependence is governed by the superconducting gap, which emerges as a sharp crossover scale that can be clearly resolved even when conventional coherence peaks are weak or absent. Thus we present bias asymmetry as an interferometric, spectroscopic probe of nonlocal interface physics and superconducting energy scales in hybrid and topological systems where extended interfaces are unavoidable.
\end{abstract}

\maketitle

Electrical transport across superconducting interfaces is most often understood in terms of Andreev reflection occurring at a sharply defined boundary between superconducting (S) and non-superconducting (N) regions\cite{Andreev1964,BTK1982}. Within this picture, quasiparticles propagate through the region N, reach the interface, and are converted into holes or electrons at opposite energies\cite{BdG,BTK1982}. Since the Bogoliubov-de Gennes (BdG) Hamiltonian possesses an exact particle-hole symmetry, this process is commonly assumed to produce conductance spectra that are symmetric with respect to bias in the Andreev regime\cite{BdG,BTK1982,Buttiker1988}. This assumption is made for a wide range of theoretical descriptions\cite{BTK1982,Lambert1998,Beenakker1997} and experimental analyses, from tunneling\cite{Rodrigo2004,Fischer2007} and point-contact spectroscopy\cite{Sheet2004,Szabo2001,Anantram1996,Samuely2009,Chen2008} to transport in semiconductor–superconductor hybrids\cite{Beenakker1992,Takayanagi1995,Mourik2012,vanWees1992,FuKane2008}. In such cases, the interface is often represented by a local scattering potential characterized by a single, bias-independent parameter e.g., as a delta-function potential barrier is assumed in the conventional Blonder-Tinkham-Klapwijk (BTK) formalism\cite{BTK1982}. In realistic devices, however, the interface typically extends over a finite spatial region. Oxide layers, depletion zones, band bending, or electrostatic confinement introduce a length scale over which quasiparticles propagate before encountering the superconducting condensate. In this regime, transport is no longer expected to be governed solely by reflection at a single point\cite{vanWees1992,Hekking1993}. While energy-dependent scattering from extended barriers is known in normal-state transport\cite{Akkermans2007}, investigation of its consequences for superconducting interfaces have remained limited\cite{Takane1992,Mortensen1999,Cuevas1996,Bozovic2002}. 

In this work we show that spatially extended superconducting interfaces make an intrinsic interferometric contribution to transport that introduces characteristic bias-asymmetry in conductance without violating the particle-hole symmetry of the BdG  Hamiltonian. The interface essentially acts as an effective Andreev interferometer with characteristic damped oscillations whose period is set by the electron-hole wave-vector mismatch\cite{Beenakker1992}. Unlike conventional Andreev interferometers based on path interference\cite{Petrashov1995,AndreevInterferometer,Bruder1993,Charlat1996}, the phase sensitivity here originates from energy-dependent propagation through the extended interface itself. 

\begin{figure*}
    \centering
    \includegraphics[width=.95\textwidth]{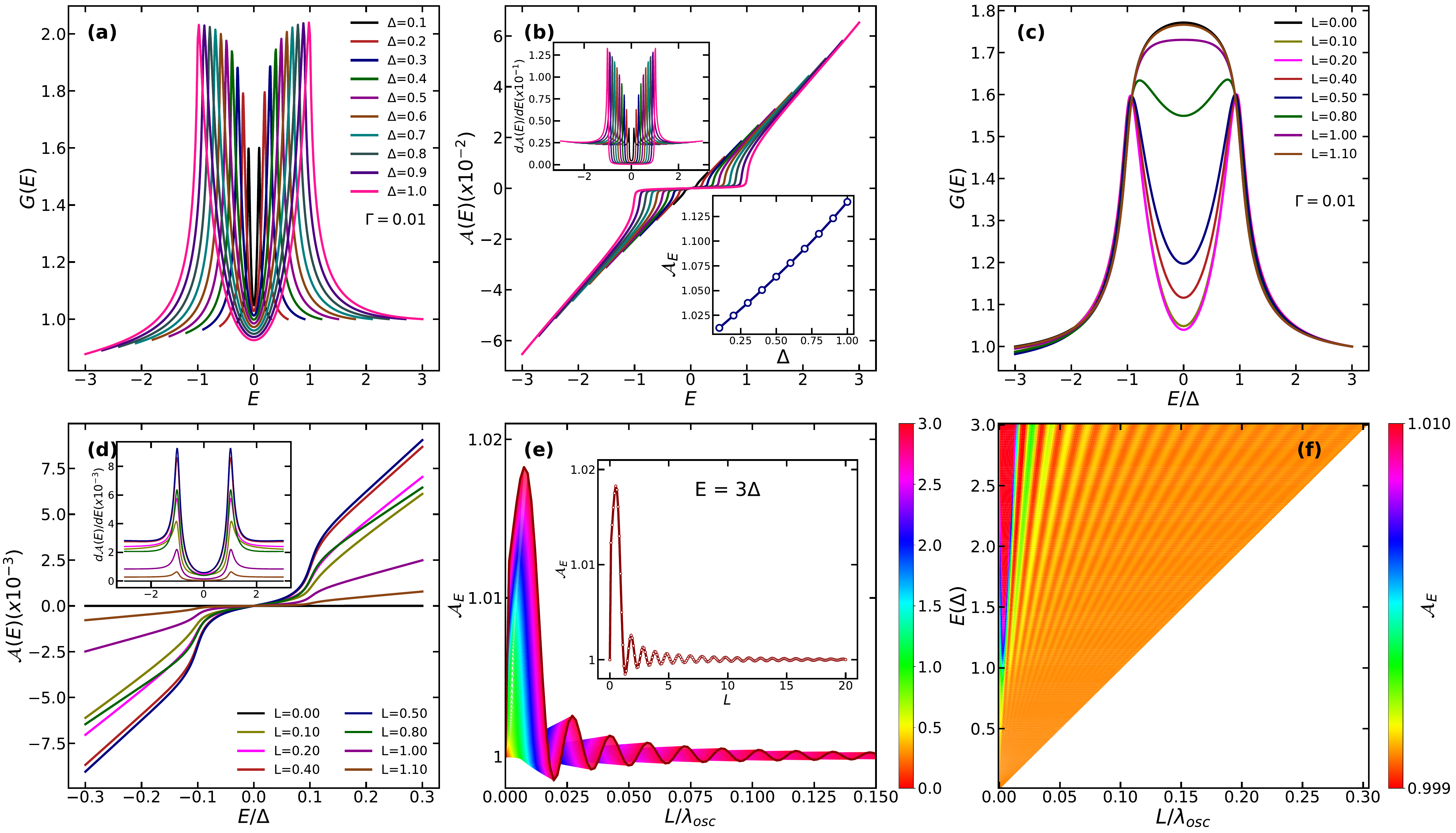}
    
    \caption{\textbf{Asymmetry and Interferometry:} 
(a) $G(E)$ vs $E$ normalized at $E=+3\Delta$ for varying $\Delta$. 
(b) $\mathcal{A}(E)$ extracted from (a) (inset upper: corresponding derivative and inset lower: $\mathcal{A}_E$ vs $\Delta$) 
(c) $G(E)$ vs $E/\Delta$ for varying $L$ normalized at $E=+3\Delta$.
(d) $\mathcal{A}(E)$ extracted from (c) (inset: corresponding derivative). 
(e) $\mathcal{A}(E)$ for varying $L$. inset: A representative trace at $E = 3\Delta$. 
(f) $\mathcal{A}_E$ vs $L/\lambda_{\mathrm{osc}}$ showing the characteristic fan-diagram and the triangular phase-space domain corresponding to propagating modes.
}
\end{figure*}

We model an NS junction using the tight-binding BdG formalism. The BdG Hamiltonian is given by
\begin{equation}
H_{\text{BdG}} =
\begin{pmatrix}
H_0 - E_F & \Delta \\
\Delta^* & -\left(H_0 - E_F\right)^*
\end{pmatrix},
\end{equation}
where, the normal state single-particle Hamiltonian is 
\begin{equation}
H_0 = -\frac{\hbar^2}{2m}\nabla^2 + V(x).
\end{equation}
Here $E_F$ is the Fermi energy, $V(x)$ represents the interface potential profile, and $\Delta$ is the superconducting pairing potential.

To implement the model numerically, we discretize the Hamiltonian using the finite difference method on a one-dimensional lattice with lattice constant $a$. This yields a nearest-neighbor tight-binding model with hopping amplitude $t = \frac{\hbar^2}{2ma^2}$. The discretized Hamiltonian can be written in terms of on-site and hopping blocks in Nambu space\cite{Lambert1998}where, 

$H_{\text{onsite}} =
\begin{pmatrix}
2t + V(x) - E_F & \Delta \\
\Delta^* & -\left(2t + V(x) - E_F\right)^*
\end{pmatrix}$,

$H_{\text{hopping}} =
\begin{pmatrix}
-t & 0 \\
0 & -t
\end{pmatrix}.$
\\

While we have considered multiple forms of $V(x)$ to establish the generic nature of the results (please see the Supplementary Information), here we present only the case of a rectangular barrier. For that,

\begin{equation}
V(x) =
\begin{cases}
V_0, & 0 \le x \le L, \\
0, & \text{otherwise}
\end{cases}
\end{equation}
with $V_0$ a constant. Following BTK, we characterize the barrier by the dimensionless barrier strength parameter $Z= \frac{V_0 L}{\hbar v_f}$, where $v_f$ is the Fermi velocity of the incoming electrons in the N region.

Transport properties are computed within the scattering-matrix formulation of the BdG problem. Details are provided in the supplemental materials. The N and S regions are modeled as semi-infinite leads attached to the interface region. The scattering problem is solved numerically using the \textsc{Kwant} package\cite{Kwant} , which allows the discretized BdG Hamiltonian to be implemented directly in tight-binding form and yields the full set of scattering amplitudes at a given quasiparticle energy. From these amplitudes, the normal and Andreev reflection probabilities for electrons incident from the N lead are extracted. In order to separate normal-state propagation through the interface region from superconducting conversion at the boundary, we introduce the normal-state scattering matrix of the interface region, denoted by \(S_N(E)\)\cite{Beenakker1997}. It is defined as the scattering matrix obtained from \(H_0\), with \(\Delta=0\). The differential conductance is then obtained from the reflection probabilities according to the standard BTK expression,
\begin{equation}
G(E)=\frac{2e^2}{h}\left[N - R_{ee}(E) + R_{he}(E)\right]_{E=eV},
\end{equation}
where \(R_{ee}(E)\) and \(R_{he}(E)\) are the probabilities for normal and Andreev reflection, respectively, and \(N\) is the number of propagating modes in the N lead\cite{Beenakker1992,Lambert1998}. All calculations are performed at zero temperature. 
In order to make the model more realistic, we also included a Dynes broadening parameter $\Gamma$\cite{Dynes1978,Plecenik1994}. This is done by evaluating the scattering matrix at complex energy $E + i\Gamma$. 

As seen in Figure 1(a), for $L \neq 0$, $G(E)$ is bias-asymmetric, for different $\Delta$. For finite $L$, quasiparticles incident from the N region must propagate over the full length of the interface region before reaching the S boundary. Doing so, an electron-like excitation at energy \(+E\) acquires a propagation factor \(\exp[i k_e(E)L]\), while a hole-like excitation at energy \(-E\) acquires \(\exp[i k_h(E)L]\), where \(k_{e,h}(E)\) are the longitudinal wave vectors in the interface region determined by the local dispersion\cite{Akkermans2007}. Depending on energy and barrier height, \(k_{e,h}(E)\) may be real, describing propagating modes and phase accumulation, or imaginary, \(k_{e,h}(E)=i\kappa_{e,h}(E)\), describing evanescent modes and exponential attenuation. Although the difference between \(k_e(E)\) and \(k_h(E)\) is small, the effect is amplified by the factor \(L\), leading to unequal accumulated phase or decay for electrons and holes. As a result, the normal-state scattering matrix becomes intrinsically asymmetric with energy, satisfying \(S_N(E)\neq S_N(-E)\). This asymmetry is directly converted into bias asymmetry by Andreev reflection, which coherently couples electron amplitudes at energy \(+E\) to hole amplitudes at energy \(-E\). This leads to \(G(+E)\neq G(-E)\). It is important to note that although the observable asymmetry in Figure 1(a,c) is modest in the parameter regime considered here, we will see later that its magnitude scales inversely with the Fermi velocity. In systems with lower carrier density, e.g., semiconductor-superconductor hybrids, the phase mismatch can be substantially enhanced, leading to a much stronger asymmetry. In fact, the asymmetry will become increasingly pronounced as the ratio $E/E_F$ grows.

To quantify the asymmetry, we define an energy dependent asymmetry factor
\begin{equation}
\mathcal{A}(E)=\frac{G(+E)-G(-E)}{G(+E)+G(-E)},
\end{equation} 
\noindent$\mathcal{A}(E)$ isolates the odd-in-bias component of the conductance. We characterize the asymmetry at a fixed $E$ as $(\mathcal{A}_E = \frac{G(+E)}{G(-E)})$. As seen in the lower inset of Figure 1(b), $\mathcal{A}_E$ for $E >\Delta$ (normal state) increases with increasing $\Delta$.

\begin{figure}
    \centering
    \includegraphics[width=.75\columnwidth]{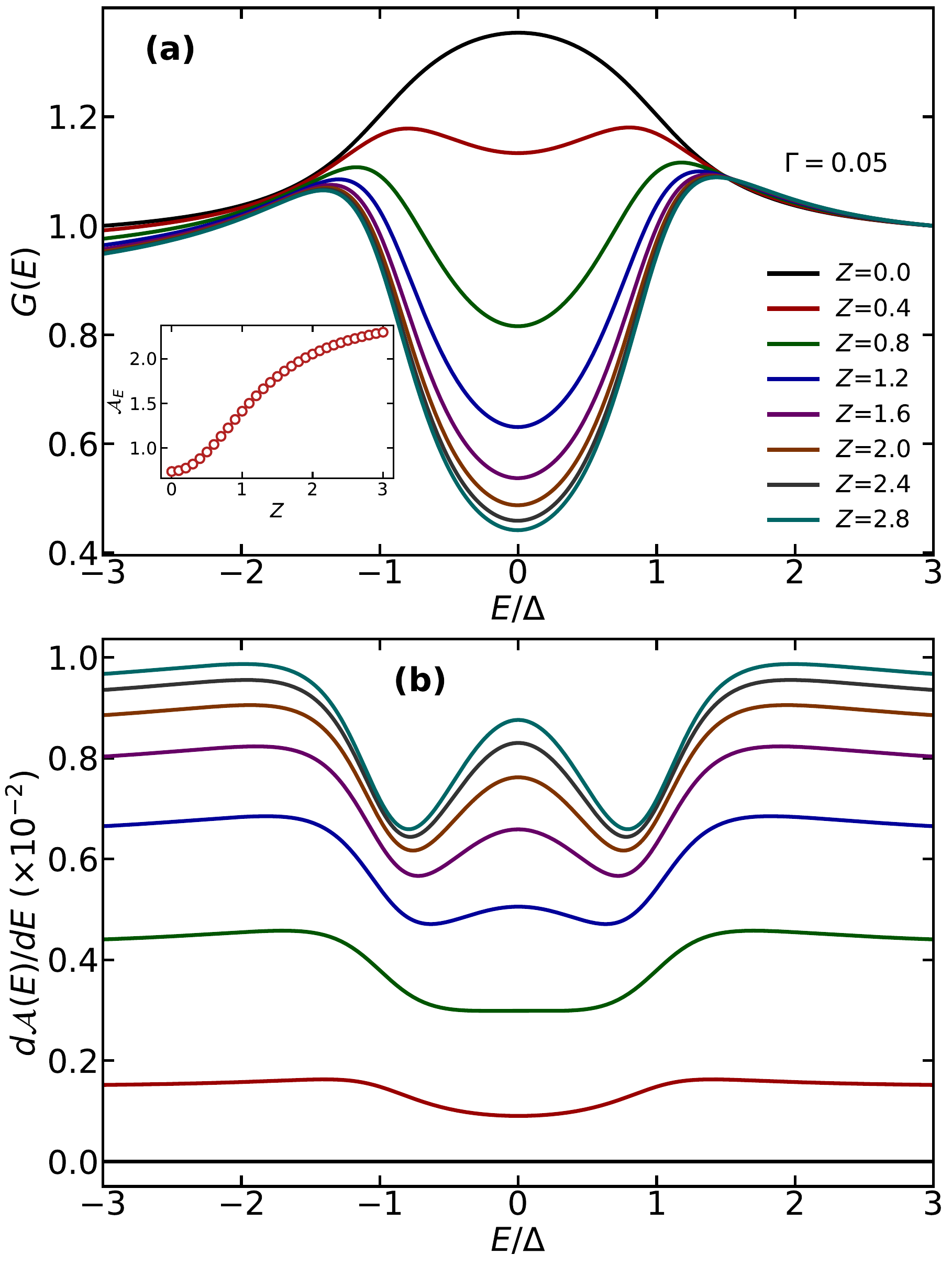}
    
    \caption{\textbf{Barrier-strength evolution:} 
(a) Normalized $G(E)$ vs $E/\Delta$ for varying $Z$ (inset: $\Delta_{\mathrm{fit}}$ vs $Z$). 
(b) Corresponding asymmetry $\mathcal{A}(E)$.}
    \label{fig:2}
\end{figure}

$G(E)$ for different $L$ and for a given $\Delta$ are shown in Figure 1(c). For $L\rightarrow 0$, the standard BTK result is reproduced with no bias-asymmetry. The inset of Figure 1(e) shows the evolution of ( representative) $\mathcal{A}_{E = 3\Delta}$ with $L$. The asymmetry initially increases with increasing $L$, reaches a maximum at an intermediate $L$, and then develops pronounced oscillations. These oscillations are superimposed on an overall decaying envelope, and eventually the asymmetry collapses for sufficiently large \(L\). The oscillatory dependence of \(\mathcal{A}_E\) on \(L\) reflects an interferometric mechanism intrinsic to Andreev transport through an extended interface. Electron and hole amplitudes that have acquired different phases within the interface region undergo coherent multiple reflections, interfering in a manner analogous to a Fabry-P\'erot cavity. Hence, the interface itself acts as an effective Andreev interferometer. Unlike conventional Andreev interferometers, which rely on spatially separated paths\cite{AndreevInterferometer}, phase-biased superconducting loops\cite{Petrashov1995,Charlat1996}, or multiple superconducting terminals\cite{Bruder1993,Anthore2003}, the interferometric effect identified here arises from coherent electron–hole propagation through a single extended interface region. The interference occurs between electron and hole amplitudes coupled at opposite energies by Andreev reflection, rather than between spatially distinct trajectories\cite{Beenakker1992,Volkov1999}.

\begin{figure}
    \centering
    \includegraphics[width=.75\columnwidth]{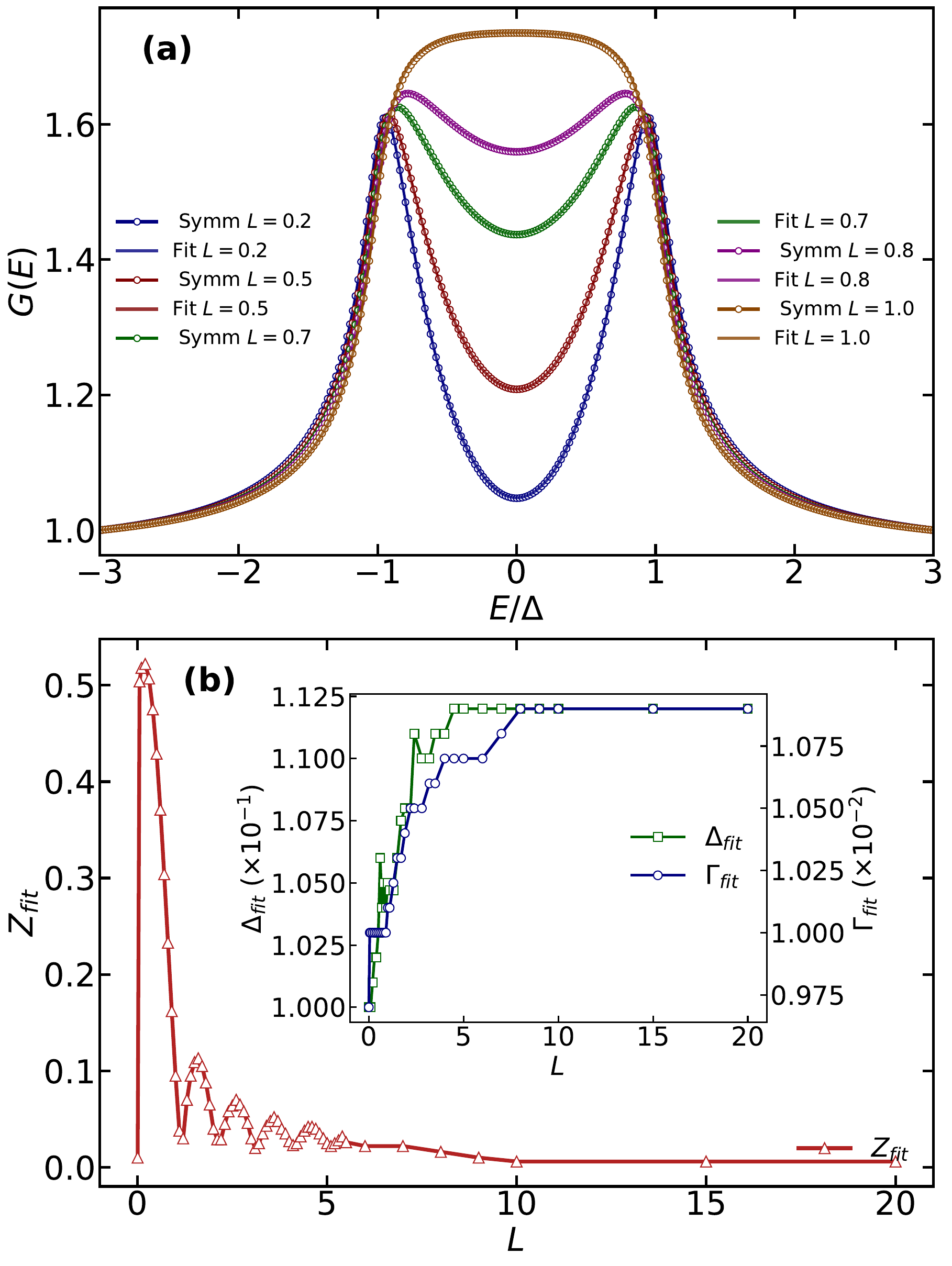}
    \caption{\textbf{Symmetrization and BTK fitting:} 
(a) Normalized $G(E)$ for different $L$: Symmetrized spectra and BTK fits. 
(b) Fitted $Z_{\mathrm{fit}}$ vs $L$; inset: $\Delta_{\mathrm{fit}}$ and $\Gamma_{\mathrm{fit}}$ vs $L$.
}
    \label{figure:3}
\end{figure}

To understand the effect quantitatively, we note that the accumulated electron-hole phase is
\begin{equation}
\Phi(E,L) = [k_e(E) - k_h(E)]L.
\end{equation}

Oscillations in the odd-in-bias\cite{note} conductance ($\mathcal{A}_E$) occur whenever
\begin{equation}
\Phi(E,L) = 2\pi n,
\end{equation}
\noindent with $n$ an integer.
This condition helps us define a characteristic oscillation length
\begin{equation}
\lambda_{\mathrm{osc}}(E)
= \frac{2\pi}{|k_e(E)-k_h(E)|},
\end{equation}
corresponding to the distance required to accumulate a $2\pi$
electron-hole phase mismatch. Now, rewriting the phase as
\begin{equation}
\Phi(E,L)=2\pi\,\frac{L}{\lambda_{\mathrm{osc}}(E)},
\end{equation}
shows that the oscillation extrema are determined by the
dimensionless variable $L/\lambda_{\mathrm{osc}}(E)$ alone.
Consequently, as shown in Figure 1(e), when $\mathcal{A}_E$ is plotted against
$L/\lambda_{\mathrm{osc}}$, the interference extrema at different
energies align at identical values of the scaled coordinate thereby making the oscillatory structure collapse onto a
single universal periodic curve. This data-collapse is a direct confirmation
that the asymmetry is indeed controlled by coherent electron-hole phase
accumulation. In the $(E,L)$ representation, the same condition generates a family of constant-phase contours which produces a characteristic fan-diagram (Figure 1(f)) indicating $\Phi \propto EL$ in the low-energy (Andreev) limit. The fan pattern occupies a triangular domain the boundary of which reflects the finite energy window over which propagating quasiparticle states exist.

Here we note that $\lambda_{\mathrm{osc}}(E)$ can be directly measured from experiments and it depends only on
$k_e(E)-k_h(E)$.
It therefore provides direct access to the electron-hole wave-vector mismatch and, through its energy dependence, yields information about Fermi velocity,
carrier density, effective mass, and band curvature within the extended interface region. 

For sufficiently large \(L\), transport is dominated by evanescent modes for which the longitudinal wave vector is purely imaginary, \(k(E)=i\kappa(E)\). In this regime, the transmission probability acquires an overall envelope that decays approximately as $T(E,L)\sim e^{-2\kappa(E)L}$. Since transport is exponentially small, the difference between electron and hole propagation also becomes experimentally irrelevant, leading to a collapse of the bias asymmetry at large $L$. Therefore, the conductance asymmetry naturally appears in three regimes as a function of $L$: (a) a short-interface regime where propagation is effectively local,(b) an intermediate regime dominated by coherent interference, and (c) an asymptotic regime where transport is evanescent and exponentially suppressed.

We now return to the bias ($E$)-dependent \(\mathcal A(E)\) as shown in Figure 1(b,d).  A striking feature of \(\mathcal A(E)\) is that for biases satisfying $|E|\lesssim\Delta$, the asymmetry exhibits only a weak and smooth $E$-dependence, whereas a rapid variation develops once $E$ approaches $\Delta$. While the asymmetric factor in experiments is usually considered a nuisance and is ignored, or is erased through symmetrization, the shape of $\mathcal{A}(E)$ for a given $L$ matches remarkably well with experiments where the antisymmetric conductance was explicitly reported\cite{Manan2010}. This behavior follows naturally from the structure of Andreev transport through an extended interface. For \(|E|<\Delta\), transport is dominated by Andreev reflection and the asymmetry is controlled primarily by \(S_N(E)\), which varies smoothly on the scale of $\Delta$. As a result, \(\mathcal A(E)\) remains weakly bias dependent throughout the Andreev regime. Once the bias exceeds $\Delta$, quasiparticle transmission into the S region becomes allowed, and the balance between Andreev reflection and normal transmission changes abruptly. This crossover is captured by the asymmetry, leading to the sharp evolution of \(\mathcal A(E)\) at \(|E|\simeq\Delta\). Therefore this feature can be used as an unconventional spectroscopic tool. Even in junctions where disorder, soft gaps, or interface inhomogeneity smear conventional coherence peaks in the differential conductance, $\Delta$ can be extracted from the bias at which \(\mathcal A(E)\) changes most rapidly. In practice, this can be made explicit by considering the derivative \(d\mathcal A/dE\), which develops clear extrema at the gap edges (see the insets of Figure 1(b,d)). For example, in Figure 1(c), for L = 1.0, the coherence peaks in $G(E)$ are absent but two clear peaks in \(d\mathcal A/dE\) reveal the gap precisely. This gap information is obtained through an interferometric response, providing complementary information to standard tunneling spectroscopy\cite{Sheet2004,Szabo2001,Samuely2009,Rodrigo2004,Fischer2007}.

In Figure 2(a) we plot $G(E)$ for a given $L$, for different $V_0$ which effectively leads to different $Z$, as in the BTK formalism. This is physically different from the $L$-dependence discussed earlier, which would also effectively lead to a change in $Z$. The asymmetry factor $\mathcal{A}(E)$ extracted from Figure 2(a) are shown in Figure 2(b). $\mathcal{A}(E)$ increases systematically with $Z$, with a zero-bias peak (ZBP) appearing and growing for higher $Z$. This ZBP is a direct consequence of interferometric phase accumulation in the tunneling regime\cite{MartinRodero2011}. As $Z$ increases, the symmetric background is strongly suppressed, while $S_N(E)$ becomes sharper. Although the electron-hole phase mismatch vanishes linearly as $E\to 0$, the relative imbalance between $S_N(E)$ and $S_N(-E)$ is most visible near zero bias when transmission is minimal but still finite (high $Z$). The resulting enhancement of the odd-in-bias component produces a zero-bias peak in $\mathcal{A}(E)$, even as the total conductance exhibits a dip (Figure 2(a)). $\mathcal A(E)$ in the normal state increases monotonically with $Z$ (inset of Figure 2(a)), demonstrating that $Z$ acts as an effective tuning parameter for the interferometric response of the interface.

It is also instructive to compare the present results with the common practice of symmetrizing conductance spectra and fitting them using the conventional BTK model\cite{BTK1982}. In Figure 3(a) we show five representative symmetrized spectra (for different $L$) with $\Delta = 0.1, \Gamma = 0.01$, along with their BTK fits. The extracted parameters are plotted in Figure 3(b). While $\Delta_{fit}$ and $\Gamma_{fit}$ are overestimated by $\sim 12\%$, $Z_{fit}$ shows pronounced oscillation with $L$. The scattering amplitude in the conventional BTK model is characterized solely by a local, energy-independent $Z_{fit}$ and hence the fitting procedure has no phase degree of freedom. Consequently, the interference-induced variations of the transmission probability are absorbed into $Z_{fit}$, which acquires an artificial oscillatory dependence on $L$. We note that $Z_{fit}$ does not correspond to a real barrier strength, in the sense that two barriers with identical $Z$ but different $(V_0, L)$ are not equivalent. This distinction highlights the failure of a single, bias-independent BTK barrier parameter to describe spectra generated by spatially extended superconducting interfaces. We note that finite-width generalizations of the BTK framework were also attempted in the past\cite{Takane1992,Mortensen1999,Cuevas1996,Bozovic2002}. However, the physics discussed here were not addressed because the resulting spectra were analyzed within an even-in-bias framework only.

In principle, the considerations above also apply to unconventional superconductors, including those with odd-parity pairing\cite{Fogelstrom1997,Covington1997,Bergeret2005}. In particular, for $p$-wave superconductors the momentum dependence of the pairing potential strongly influences Andreev reflection, leading to directional effects and, in many cases, surface Andreev bound states\cite{Tanaka1995,Kashiwaya2000}. These features primarily modify the even-in-bias structure, and do not generate conductance asymmetry. As a result, bias asymmetry in junctions involving $p$-wave or effectively $p$-wave superconductors does provide information about interface structure, and can coexist with symmetric zero-energy features such as Majorana bound states\cite{Kitaev2001,Alicea2012,FuKane2008,Mourik2012}. Experimentally, in many hybrid superconducting devices, the effective spatial extent of the interface can be tuned electrostatically through gate-controlled depletion and band bending\cite{Mourik2012,Takayanagi1995,Churchill2013}, providing a natural platform to probe the oscillations. In more conventional planar junctions and point-contact geometries, variations in oxide thickness, interface quality, or fabrication conditions can set different regimes of interface length, making comparative studies across multiple devices possible. 

In conclusion, our results present extended superconducting interfaces as active phase-sensitive quantum elements. As hybrid and topological superconducting devices often rely on soft, electrostatically defined, or structurally imperfect interfaces, the interferometric effects identified here are likely to be unavoidable. The fundamental mechanism presented here is not limited to one-dimensionality and/or specific barrier profile. While quantitative details like the oscillation amplitude may be modified in multichannel or weakly disordered junctions\cite{Beenakker1997,Lambert1998}, the phase-sensitive origin of the odd-in-bias response remains valid. While bias-asymmetry is traditionally ignored, our work suggests that it can be harnessed as a phase-sensitive spectroscopic probe of interface structure and pairing physics. The results, therefore, open the door to exploiting bias asymmetry to characterize unconventional superconductors, multiband systems, and engineered superconducting heterostructures, and to using extended interfaces themselves as controllable building blocks for Andreev interferometry in low-dimensional quantum devices.

The authors thank Dr. Deepti Rana for critically reading the manuscript and for sharing valuable comments. The materials that support the findings of this study are available within the article and the supplementary file. The numerical codes used for the simulations are available from the corresponding author upon reasonable request. Additional information/materials will also be shared upon reasonable requests.  

\begin{thebibliography}{99}
\bibitem{note} Under $E \to -E$, particle--hole symmetry implies 
$k_e(-E) = k_h(E)$ and $k_h(-E) = k_e(E)$, so that 
$\Phi(-E,L) = -\Phi(E,L)$. 
Interference terms proportional to $\sin(2\Phi)$ therefore change sign 
under $E \to -E$ and contribute exclusively to the odd-in-bias component 
of the conductance, while $\cos(2\Phi)$ terms remain even. 
The observed asymmetry thus directly isolates the phase-sensitive part 
of the interference.
\end{thebibliography}

\end{document}